\documentclass[fleqn,usenatbib,useAMS]{mnras}

\usepackage{graphicx}	
\usepackage{amsmath}	
\usepackage{multicol}        
\usepackage{bm}		
\usepackage{pdflscape}	
\usepackage{latexsym}
\usepackage[authoryear]{natbib}
\usepackage{bm,nicefrac}

\usepackage[T1]{fontenc}
\usepackage{ae,aecompl}

\usepackage{newtxtext,newtxmath}

\def\deg{\ifmmode^\circ\else$^\circ$\fi}

\def\Q{\ifmmode\mathcal{Q}\else$\mathcal{Q}$\fi}
\def\Mach{\ifmmode\mathcal{M}\else$\mathcal{M}$\fi}

\title[colliding filamentary clouds in IRDC G333.73+0.37]
{New evidences in IRDC G333.73+0.37: colliding filamentary clouds, hub-filament system, and embedded cores}

\author[L.~K. Dewangan]
{L.~K. Dewangan$^{1}$\thanks{lokeshd@prl.res.in}\\
$^{1}$Physical Research Laboratory, Navrangpura, Ahmedabad - 380 009, India.\\}

\begin{document}

\date{ }

\pagerange{\pageref{firstpage}--\pageref{lastpage}} \pubyear{2020}

\maketitle

\label{firstpage}

\begin{abstract}
To unravel the star formation process, we present a multi-scale and multi-wavelength study of the filamentary infrared dark cloud (IRDC) G333.73+0.37, which hosts previously known two H\,{\sc ii} regions located at its center. Each H\,{\sc ii} region is associated with a mid-infrared source, 
and is excited by a massive OB star. Two filamentary structures and a hub-filament system (HFS) associated with one H\,{\sc ii} region are investigated in absorption using the {\it Spitzer} 8.0 $\mu$m image. The $^{13}$CO(J = 2--1) and C$^{18}$O(J = 2--1) line data reveal two velocity components (around $-$35.5 and $-$33.5 km s$^{-1}$) toward the IRDC, favouring the presence of two filamentary clouds at different velocities. 
Nonthermal (or turbulent) motions are depicted in the IRDC using the C$^{18}$O line data. 
The spatial distribution of young stellar objects (YSOs) identified using the VVV near-infrared data traces star formation activities in the IRDC. 
Low-mass cores are identified toward both the H\,{\sc ii} regions using the ALMA 1.38 mm continuum map. 
The VLT/NACO adaptive-optics L$^{\prime}$-band images show the presence of at least three point-like sources and the absence of small-scale features in the inner 4000 AU around YSOs NIR31 and MIR 16 located toward the H\,{\sc ii} regions. 
The H\,{\sc ii} regions and groups of YSO are observed toward the central part of the IRDC, where the two filamentary clouds intersect. 
A scenario of cloud-cloud collision or converging flows in the IRDC seems to be applicable, which may explain star formation activities including HFS and massive stars.
\end{abstract}
%
\section{Introduction}
\label{sec:intro}
In recent years, with the availability of several ground- and space-based telescopes operating in infrared, sub-millimeter (mm), and radio wavelengths, significant progress has been made by researchers to probe the physical processes involved in forming massive stars \citep[$\gtrsim$ 8 M$_{\odot}$; e.g.,][]{zinnecker07,tan14,motte18}. However, the birth of massive stars is not yet completely understood. The popular physical processes of massive star formation (MSF) reported in the literature are the cloud-cloud collision \citep[CCC;][]{habe92} and the global non-isotropic collapse (GNIC) scenario \citep{tige_2017,motte18}. The former process is closely related to the physical scenario of converging gas flows \citep{ballesteros99,vazqez07,heitsch08}. The latter process includes the ingredients of the competitive accretion (CA) model \citep{bonnell_2001,bonnell_2004,bonnell_2006} and the global hierarchical collapse (GHC) model \citep{vazquez_2009,vazquez_2017,vazquez_2019}. By taking a holistic view of massive star-forming regions, the filamentary web structures or hub-filament systems (HFSs) -- where a central hub is surrounded by  filaments -- are often encountered by researchers. Massive stars and stellar clusters are exclusively found toward the central hubs. The observed HFS may reflect an imprint of the onset of the GNIC scenario \citep{motte18}, and may have originated by collision of clouds \citep{balfour15}. To carefully assess the existing theoretical models of MSF, the knowledge of the cloud to core scale physical environments of promising massive star-forming regions is crucial. In this context, this paper examines the large- and small-scale environments of an Infrared Dark 
Cloud (IRDC) G333.73+0.37 (hereafter, G333.73).

G333.73+0.37 shows the absorption features against the Galactic background in the {\it Spitzer} images at 3.6 -- 8.0 $\mu$m images, and 
has a filamentary appearance \citep{sanchez13,veena18}. In the sub-mm and mm images, G333.73 has also elongated filamentary morphology in emission \citep{beltran06,veena18}, which is identified as a supercritical filament \citep{veena18}. 
\citet{veena18} computed the total mass of the IRDC to be $\sim$4700 M$_{\odot}$ assuming a distance of 2.6 kpc, which contains several dust clumps \citep[having temperature $\sim$14.3 -- 22.3~K; mass $\sim$87 -- 1530 M$_{\odot}$;][]{veena18}. 
Two bright infrared sources S1 and S2 associated with H\,{\sc ii} regions are embedded in G333.73 \citep[see Figures 1 and 18 in][]{veena18}, and are clearly evident in the mid-infrared (MIR) image at 8.0 $\mu$m. 
The source S1 is associated with IRAS 16164-4929 or the MIR bubble MWP1G333726+003642 \citep[see][]{veena18}. 
In the radio continuum maps at 610--1300 MHz, they detected the shell-like and compact morphology of H\,{\sc ii} regions associated with S1 and S2, respectively, which were excited by late O or early B type ZAMS stars. One point-like source MIR 16 toward S1 was found at the center of the radio continuum emission at 1300 MHz, and another source NIR31 was proposed as an exciting candidate of the H\,{\sc ii} region associated with S2 \citep{veena18}. They also referred to the source S2 as NIR31. Noticeable star formation activities (including H\,{\sc ii} regions) were reported toward the G333.73 filament \citep[see Figure 16 in][]{veena18}. Both the sources NIR31 and MIR 16 were classified as 
young stellar objects (YSOs). They studied different molecular species (HCO$^{+}$, H$^{13}$CO$^{+}$, HCN, HNC, N$_{2}$H$^{+}$, C$_{2}$H, $^{12}$CO and $^{13}$CO) in the G333.73 filament. Such study allowed them to infer signatures of accretion flows along the filament and protostellar infall toward S1. 

Based on the distribution of the dust continuum clumps at 870 $\mu$m \citep[from][]{urquhart18} and their estimated distances, we adopt a distance of 2.35 kpc to the G333.73 filament in this paper (see Section~\ref{ssec:ss1} for more details). 
Considering the previously reported findings in the target site, we find that the {\it Spitzer} images at 3.6 -- 8.0 $\mu$m (resolution $\sim$2$''$) and the molecular line data are not carefully examined to study the inner structures of the G333.73 filament hosting S1 and S2. Furthermore, cores hosting embedded YSOs (i.e., MIR 16 and NIR31) and their inner environments are yet to be studied in the G333.73 filament. Overall, in the literature, no physical process has been discussed to explain the birth of massive stars in the G333.73 filament. 

In this paper, we employed multi-scale and multi-wavelength observations to study the physical processes operating in the G333.73 filament. In particular, high resolution near-infrared (NIR) images (including adaptive-optics data; resolution $\sim$0\farcs1--0\farcs8), the ALMA 1.38 mm continuum map (resolution $\sim$0\farcs8), and molecular (i.e., $^{13}$CO(J = 2--1) and C$^{18}$O(J = 2--1)) line data are utilized to examine the large- and small-scale environment of the filament including S1 and S2. 

The structure of this paper is as follows: in Section~\ref{sec:obser}, we start with the description of data sets used in this work. In Section~\ref{sec:data}, we elaborate the observational results extracted in this paper. In Section~\ref{sec:disc}, we discuss the interpretation of the derived results. Finally, in Section~\ref{sec:conc}, we give a summary of the major findings.
\section{Data sets and analysis}
\label{sec:obser}
This paper employed several observational data sets from different existing surveys (see Table~\ref{tab1}), which were collected 
for an area of $\sim$25$'$.5 $\times$ 18$'$.54 (central coordinates: {\it l} = 333$\degr$.715; {\it b} = 0$\degr$.3507 ; see Figure~\ref{fig1}) containing G333.73.

Additionally, the {\it Herschel} temperature ($T_\mathrm{d}$) and column density maps \citep[resolution $\sim$12$''$;][]{molinari10b,marsh15,marsh17} of G333.73 were utilized in this paper. 
In the direction of S1 and S2, the primary-beam corrected ALMA continuum maps at 1.38 mm (resolution $\sim$0$''$.8 $\times$ 0$''$.77, P.A. = $-$37$\degr$.5 -- $-$44$\degr$.9) are available, and were obtained from the ALMA science archive (project \#2016.1.00191.S; PI: S{\'a}nchez-Monge, {\'A}.). We also downloaded the NAOS-CONICA (NACO) adaptive-optics L$^{\prime}$ images (resolution $\sim$0\farcs1) toward S1 and S2 (ESO proposal ID: 091.C-0379(A); PI: Jo\~{a}o Alves). In this connection, the 8.2m VLT with NACO adaptive-optics system \citep{lenzen03,rousset03} was employed. The raw NACO imaging data were processed in a similar way as outlined in \citet{dewangan15,dewangan16b,dewangan18}. 
 
The reliable photometric magnitudes of point-like sources at VVV HK$_{s}$ bands were extracted from the VVV DR4 catalog. 
Bright sources are saturated in the VVV survey \citep{minniti10,saito12,minniti17}. 
Hence, 2MASS photometric data were adopted for the bright sources. The positions of the ATLASGAL 870 $\mu$m dust continuum clumps \citep[from][]{urquhart18} were also collected in our target selected area.

To study the spatial-kinematic structure of the G333.73 filament, the $^{13}$CO(J =2$-$1) and C$^{18}$O(J =2$-$1) line data were collected from the Structure, Excitation and Dynamics of the Inner Galactic Interstellar Medium \citep[SEDIGISM;][]{schuller17} survey. 
To improve sensitivities, we smoothed these SEDIGISM data cubes with a Gaussian function with a width of 3 pixels.
\section{Results}
\label{sec:data}
\subsection{Physical environment of the G333.73 filament}
\label{ssec:ss1}
\begin{figure*}
\includegraphics[width=11.2cm]{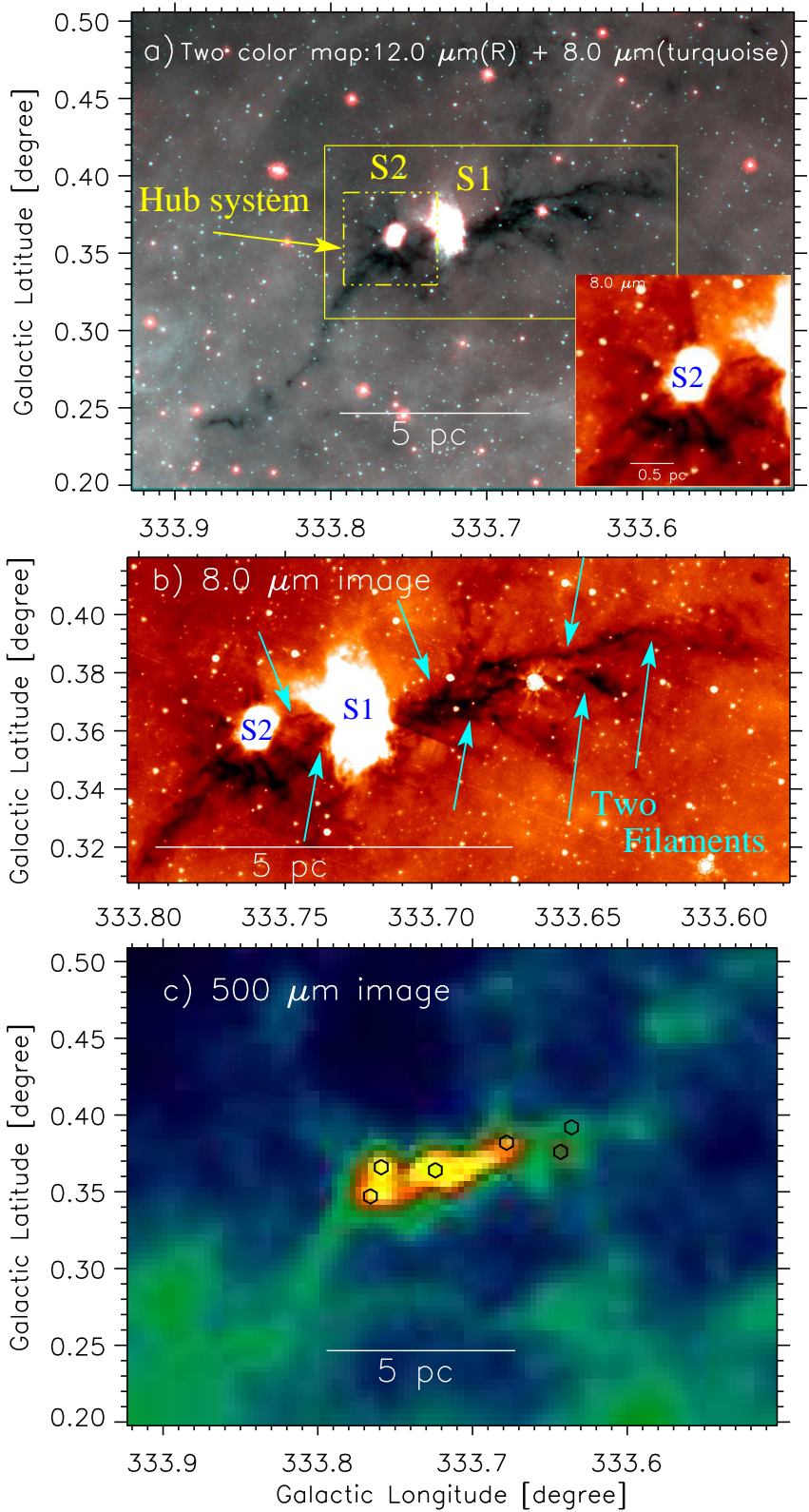}
\caption{a) The panel shows a two color-composite map of the filamentary IRDC G333.73+0.37 
(area $\sim$25\rlap.{$'$}5 $\times$ 18\rlap.{$'$}54 centered at {\it l} = 333$\degr$.715; {\it b} = 0$\degr$.351). 
The color-composite map is produced using the WISE 12.0 $\mu$m (red) and {\it Spitzer} 8.0 $\mu$m (turquoise) images. 
Two distinct MIR sources S1 and S2 are labeled in the panel. 
Using the {\it Spitzer} 8.0 $\mu$m image, the inset on the bottom right shows the region around S2 in 
zoomed-in view (see a dotted-dashed box in Figure~\ref{fig1}a), revealing a HFS toward S2. 
The solid box (in yellow) encompasses the area shown in Figure~\ref{fig1}b. 
b) The panel presents the {\it Spitzer} 8.0 $\mu$m image of an area highlighted by a solid box in Figure~\ref{fig1}a. 
At least two filamentary features are indicated by curves. 
c) The panel displays the {\it Herschel} 500 $\mu$m image of IRDC G333.73+0.37. 
The positions of the ATLASGAL dust continuum clumps at 870 $\mu$m \citep[from][]{urquhart18} 
are also marked by hexagons. 
The scale bar corresponding to 5 pc (at a distance of 2.35 kpc) is shown in each panel.} 
\label{fig1}
\end{figure*}

In Figure~\ref{fig1}a, we present a two color-composite map of the G333.73 filament using the WISE 12.0 $\mu$m (in red) and {\it Spitzer} 8.0 $\mu$m (in turquoise) images. In both the MIR images, the absorption features against the Galactic background clearly trace the IRDC, which hosts the MIR sources S1 and S2 associated with the bright extended emissions. 
Our careful inspection of the MIR images reveals that the MIR source S2 is surrounded by several filaments in the  absorption, revealing a HFS (see the inset in Figure~\ref{fig1}a). Furthermore, using the {\it Spitzer} 8.0 $\mu$m image, we find the presence of two 
filamentary features (length $>$ 6 pc) in absorption (see arrows in 
Figure~\ref{fig1}b). These are the new findings in the G333.73 filament.

Figure~\ref{fig1}c displays the {\it Herschel} 500 $\mu$m continuum image (resolution $\sim$37$''$), showing the G333.73 filament in emission.
Note that the resolution of this {\it Herschel} sub-mm image is not enough to depict the two 
filamentary features seen in the {\it Spitzer} 8.0 $\mu$m image.
We find the positions of six ATLASGAL 870 $\mu$m dust continuum clumps toward the G333.73 filament (see hexagons in Figure~\ref{fig1}c), which are located at a distance of 2.35 kpc and are traced in a velocity range of [$-$35.7, $-$33] km s$^{-1}$ \citep[from][]{urquhart18}.

\begin{figure*}
\includegraphics[width=13.3cm]{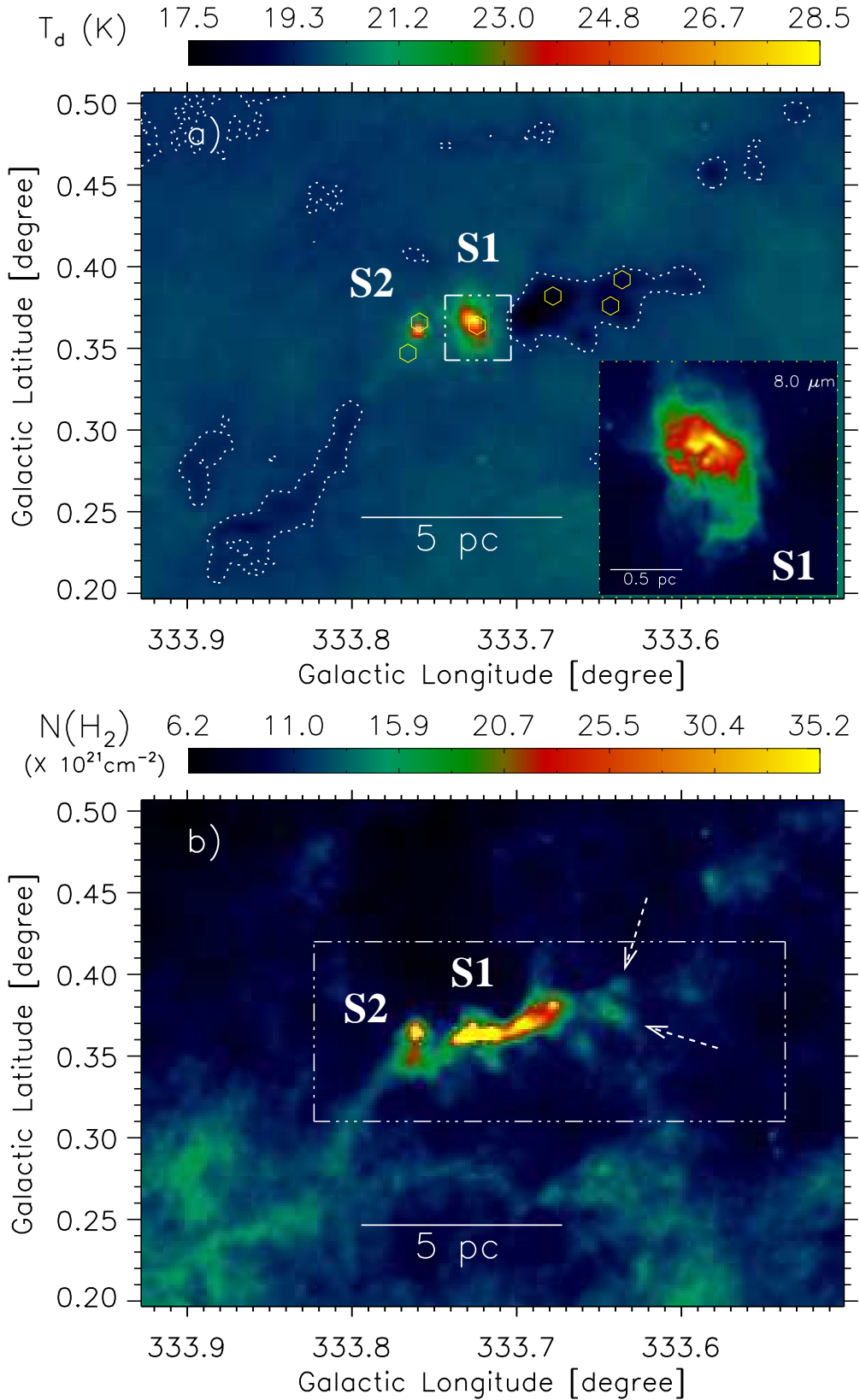}
\caption{a) {\it Herschel} temperature map (T$_{d}$) toward IRDC G333.73+0.37. 
A dotted contour at T$_\mathrm{d}$ $\sim$19.5 K is also shown in the panel. 
Using the {\it Spitzer} 8.0 $\mu$m image, the inset on the bottom right shows the region around S1 in 
zoomed-in view (see a dotted-dashed box in Figure~\ref{fig2}a). 
The positions of the ATLASGAL dust continuum clumps at 870 $\mu$m \citep[from][]{urquhart18} are also marked by hexagons. 
b) {\it Herschel} column density map toward IRDC G333.73+0.37. 
The scale bar corresponding to 5 pc (at a distance of 2.35 kpc) is shown in each panel.} 
\label{fig2}
\end{figure*}

Figures~\ref{fig2}a and~\ref{fig2}b show the {\it Herschel} temperature and column density ($N(\mathrm H_2)$) maps (resolution $\sim$12$''$) of the G333.73 filament, respectively. The inset in Figure~\ref{fig2}a also presents a zoomed-in view of S1 using the {\it Spitzer} 8.0 $\mu$m image, which is found to be associated with the radio continuum emission. A point-like source is seen in the inset, and has been identified as a Class~I YSO labeled as MIR 16 by \citet{veena18}. A contour at T$_\mathrm{d}$ $\sim$19.5~K is also shown in the {\it Herschel} temperature map (see Figure~\ref{fig2}a). We notice the temperature gradient in the filament.  In Figure~\ref{fig2}a, the areas around the MIR sources S1 and S2 (or previously known H\,{\sc ii} regions) appear to be associated with warm dust emission (i.e. 24--28~K), while other parts of the filament show a relatively cold dust emission (i.e. 17.5--20~K). 
A HFS associated with S2 is not resolved in the {\it Herschel} column density map, but two arrows are marked to highlight two filamentary features seen in the {\it Spitzer} 8.0 $\mu$m image (see Figure~\ref{fig2}b). 
\subsection{Distribution of molecular gas in G333.73}
\label{ssec:mmap}
\begin{figure*}
\includegraphics[width=\textwidth]{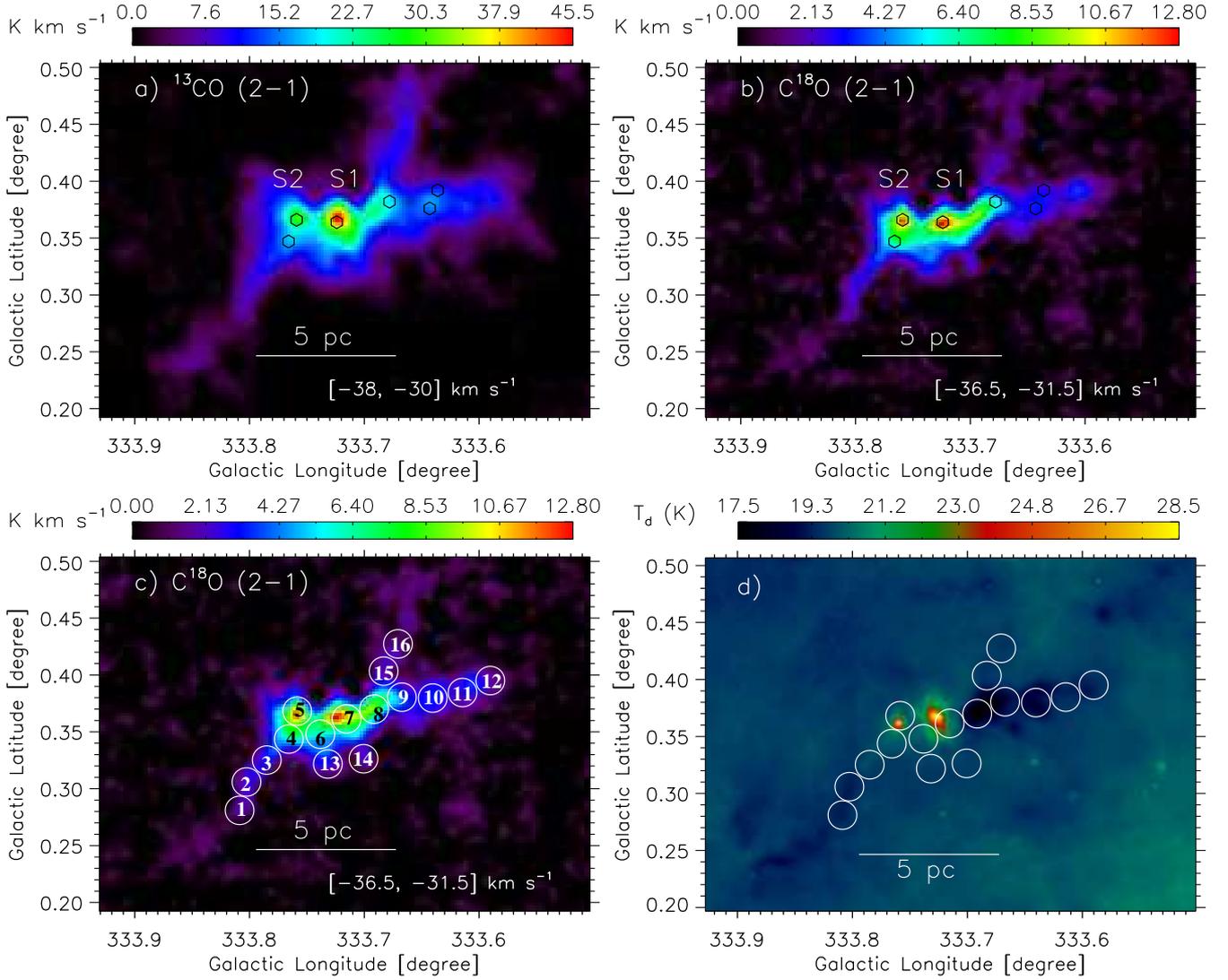}
\caption{SEDIGISM integrated intensity (or moment-0) maps of a) $^{13}$CO(J =2$-$1) and b) C$^{18}$O(J =2$-$1) 
toward IRDC G333.73+0.37. 
The molecular emission is integrated over a velocity interval, which is given in each panel (in km s$^{-1}$).
c) same as Figure~\ref{fig3}b, but overlaid with 16 circular regions (radius = 45$''$) distributed 
along the filamentary structure. d) Overlay of 16 circular regions (see Figure~\ref{fig3}c) on the {\it Herschel} temperature map.
In panels ``a" and ``b", the symbols are the same as in Figure~\ref{fig1}c.}
\label{fig3}
\end{figure*}

\begin{figure*}
\includegraphics[width=\textwidth]{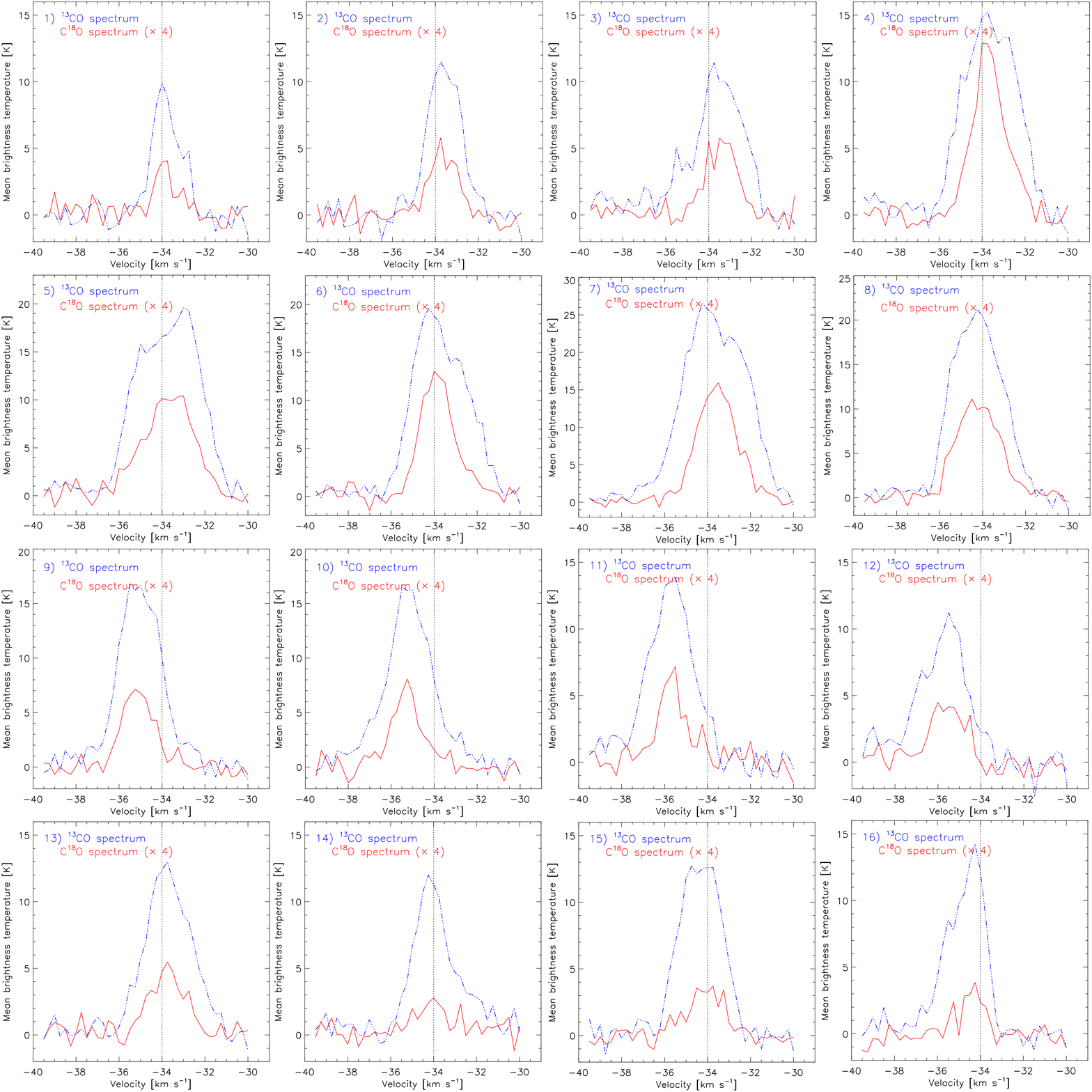}
\caption{Spectra of the $^{13}$CO(J =2$-$1) and C$^{18}$O(J =2$-$1) emission toward 16 circular regions 
marked in Figure~\ref{fig3}c. The corresponding circle number is indicated in each panel. 
A vertical dotted line in each panel highlights the velocity at $-$34 km s$^{-1}$.} 
\label{fig4}
\end{figure*}

\begin{figure*}
\includegraphics[width=\textwidth]{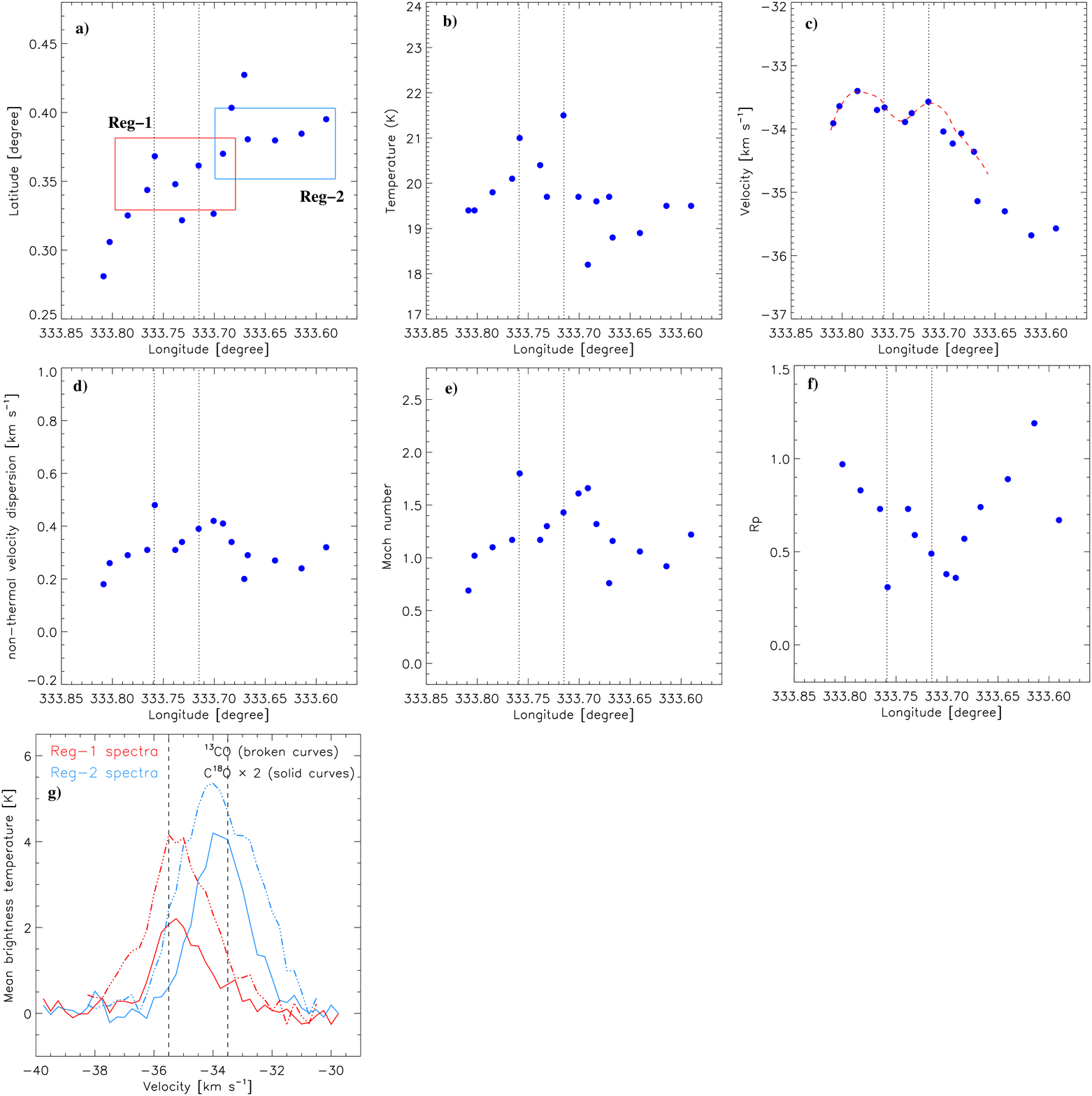}
\caption{a) The panel shows the positions of 16 selected regions (see big circles in Figure~\ref{fig3}c). 
Two regions Reg-1 and Reg-2 are also highlighted by rectangles, where average spectra are produced (see Figure~\ref{fig5}g). 
b--f) Distribution of the dust temperature, radial velocity, non-thermal velocity dispersion, Mach number, 
ratio of thermal to non-thermal gas pressure toward the selected regions against the Galactic longitude. 
The dust temperature is determined from the {\it Herschel} temperature map, while other physical parameters 
are derived using the C$^{18}$O line data. g) Spectra of the $^{13}$CO(J =2$-$1) and C$^{18}$O(J =2$-$1) emission toward Reg-1 and Reg-2 
(see solid boxes in Figure~\ref{fig5}a). In all panels (except ``g"), dotted lines show the locations of the MIR sources S1 and S2.} 
\label{fig5}
\end{figure*}

In this section, we examine the spatial and velocity distribution of the SEDIGISM $^{13}$CO(J = 2--1) and C$^{18}$O(J = 2--1) emission. 
Figure~\ref{fig3}a displays the $^{13}$CO integrated map at [$-$38, $-$30] km s$^{-1}$, 
while the C$^{18}$O emission map at [$-$36.5, $-$31.5] km s$^{-1}$ is presented in Figure~\ref{fig3}b. Both the molecular maps are overlaid with the positions of the ATLASGAL clumps. The filamentary morphology is seen in both the molecular maps, and its central part hosting S1 and S2 is depicted with high intensities. 
The C$^{18}$O emission is known to trace the denser regions compared to the $^{13}$CO emission. 
We have selected 16 circular regions (radius = 45$''$) distributed along the filamentary structure, which are indicated in the C$^{18}$O integrated map (see Figure~\ref{fig3}c) and the {\it Herschel} temperature map (see Figure~\ref{fig3}d). Five circular regions (\#4--8) are located toward the central part of the filament. 
Average spectra of $^{13}$CO(J = 2--1) and C$^{18}$O(J = 2--1) toward each circular region are presented in Figure~\ref{fig4}, allowing us to infer the peak velocity and the measured Full Width Half Maximum (FWHM) line width ($\Delta V$) of each observed spectrum. Using the C$^{18}$O line data, we have computed the non-thermal velocity dispersion, sound speed, Mach number, ratio of thermal to non-thermal pressure toward each circular region. Average dust temperature is also determined for each circular region using the {\it Herschel} temperature map. All these parameters are listed in Table~\ref{tab2} (see also Figure~\ref{fig5}).
Mach number refers to the ratio of non-thermal velocity dispersion ($\sigma_{\rm NT}$) to sound speed ($a_{s}$). 
The sound speed $a_{s}$ (= $(k T_{kin}/\mu m_{H})^{1/2}$) can be estimated using the value of gas kinetic temperature (T$_{kin}$) and mean molecular weight ($\mu$=2.37; approximately 70\% H and 28\% He by mass).
Following the work of \citet{lada03}, the ratio of thermal to non-thermal (or turbulent) pressure (i.e., $R_{p} = {a_s^2}/{\sigma^2_{NT}}$) is also determined. The following equation is used to obtain the non-thermal velocity dispersion:
\begin{equation}
\sigma_{\rm NT} = \sqrt{\frac{\Delta V^2}{8\ln 2}-\frac{k T_{kin}}{29 m_H}} = \sqrt{\frac{\Delta V^2}{8\ln 2}-\sigma_{\rm T}^{2}} ,
\label{sigmanonthermal}
\end{equation}
where $\Delta V$ is defined earlier, $\sigma_{\rm T}$ (= $(k T_{kin}/30 m_H)^{1/2}$) is the thermal broadening for C$^{18}$O at T$_{kin}$, and $m_H$ is the mass of hydrogen atom. Here, we adopt the average dust temperature for each circular region. 

Figure~\ref{fig5}a shows the locations of all the circular regions. In Figures~\ref{fig5}b,~\ref{fig5}c,~\ref{fig5}d,~\ref{fig5}e, and~\ref{fig5}f, we present the distribution of the dust temperature, peak velocity, $\sigma_{\rm NT}$, Mach number, and $R_{p}$ against the the Galactic longitude, respectively.
There is a hint of velocity variation along the filament, which is indicated by a broken curve in Figure~\ref{fig5}c. We find that Mach number is larger than one for most of the selected circular regions (see Figure~\ref{fig5}e). From Figure~\ref{fig5}f, we find that R$_{p}$ is smaller than one for all the selected regions (except one circular region), suggesting that the non-thermal pressure exceeds the thermal pressure in the filament. 

Figure~\ref{fig5}g displays the spectra of $^{13}$CO(J =2$-$1) and C$^{18}$O(J =2$-$1) toward two regions Reg-1 and Reg-2, which are indicated by solid boxes in Figure~\ref{fig5}a. The $^{13}$CO and C$^{18}$O spectra peak around $-$35.5 km s$^{-1}$ in the direction of Reg-1, while in the direction of Reg-2, 
we find the peaks of the molecular spectra around $-$33.5 km s$^{-1}$. Five circular regions (\#4--8) are located toward the Reg-1, while 
five circular regions (\#8--12) are found in the direction of Reg-2. From Figure~\ref{fig5}g, we find a significant velocity variation in the filament, indicating the presence of two velocity components in the direction of the filament. 

\begin{figure*}
\includegraphics[width=\textwidth]{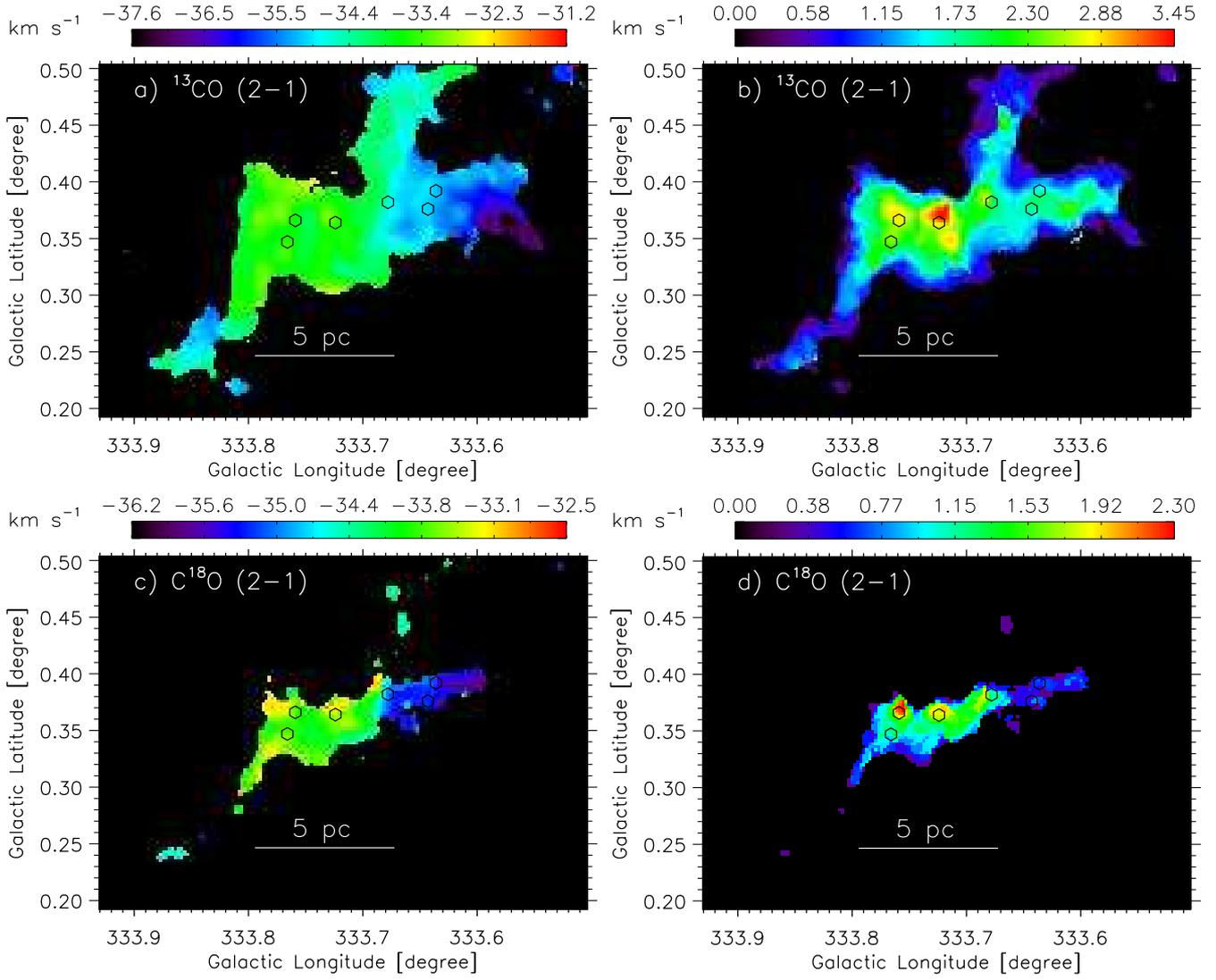}
\caption{SEDIGISM moment-1 map of a) $^{13}$CO(J =2$-$1) and c) C$^{18}$O(J =2$-$1) toward IRDC G333.73+0.37. 
SEDIGISM moment-2 map of b) $^{13}$CO(J =2$-$1) and d) C$^{18}$O(J =2$-$1) toward IRDC G333.73+0.37. 
In each panel, the symbols and the scale bar are the same as in Figure~\ref{fig1}c.} 
\label{fig6}
\end{figure*}

In Figures~\ref{fig6}a and~\ref{fig6}b, we display the line velocity/velocity field/moment-1 map and 
the intensity-weighted line width map (moment-2) of $^{13}$CO(J = 2--1), respectively.
The moment-1 and moment-2 maps of C$^{18}$O(J = 2--1) are presented in Figures~\ref{fig6}c and~\ref{fig6}d, respectively.
In all the panels of Figure~\ref{fig6}, the positions of the ATLASGAL clumps are also marked. 
Noticeable velocity gradient toward the filament is clearly depicted in both the moment-1 maps. The central part of the filament hosting S1 and S2 shows higher values of line widths ($>$ 1.5 km s$^{-1}$). 

\begin{figure*}
\includegraphics[width=15cm]{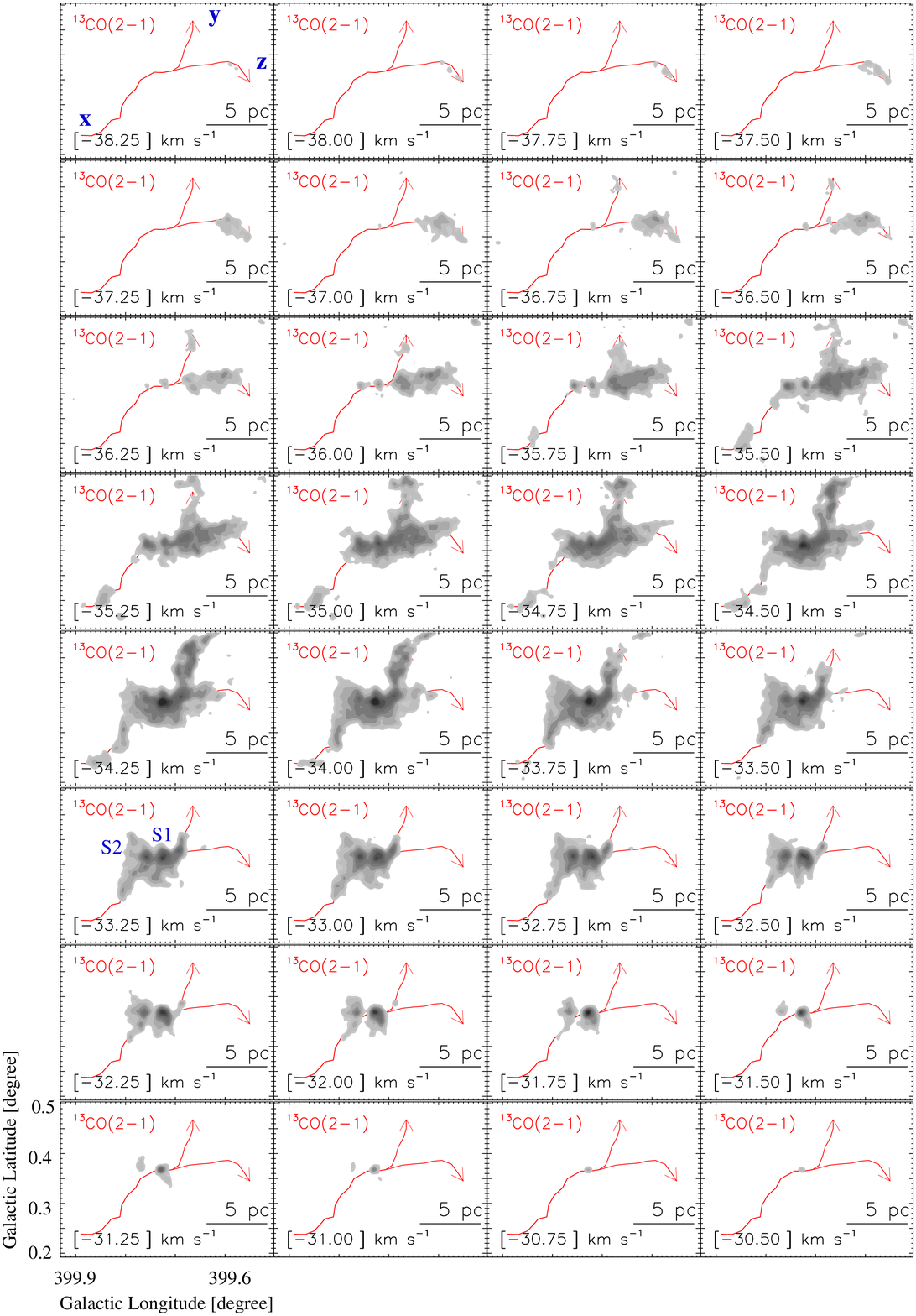}
\caption{Velocity channel maps of the SEDIGISM $^{13}$CO(2--1) emission. 
The velocity value (in km s$^{-1}$) is labeled in each panel. 
The contour levels are 1.5, 2, 3, 4, 5, 6, 7, 8, 9, and 10 K. 
Two curves (i.e., ``xy" and ``xz") are also indicated in each panel. 
In each panel, the scale bar is the same as in Figure~\ref{fig1}a.}
\label{fig7}
\end{figure*}

\begin{figure*}
\includegraphics[width=\textwidth]{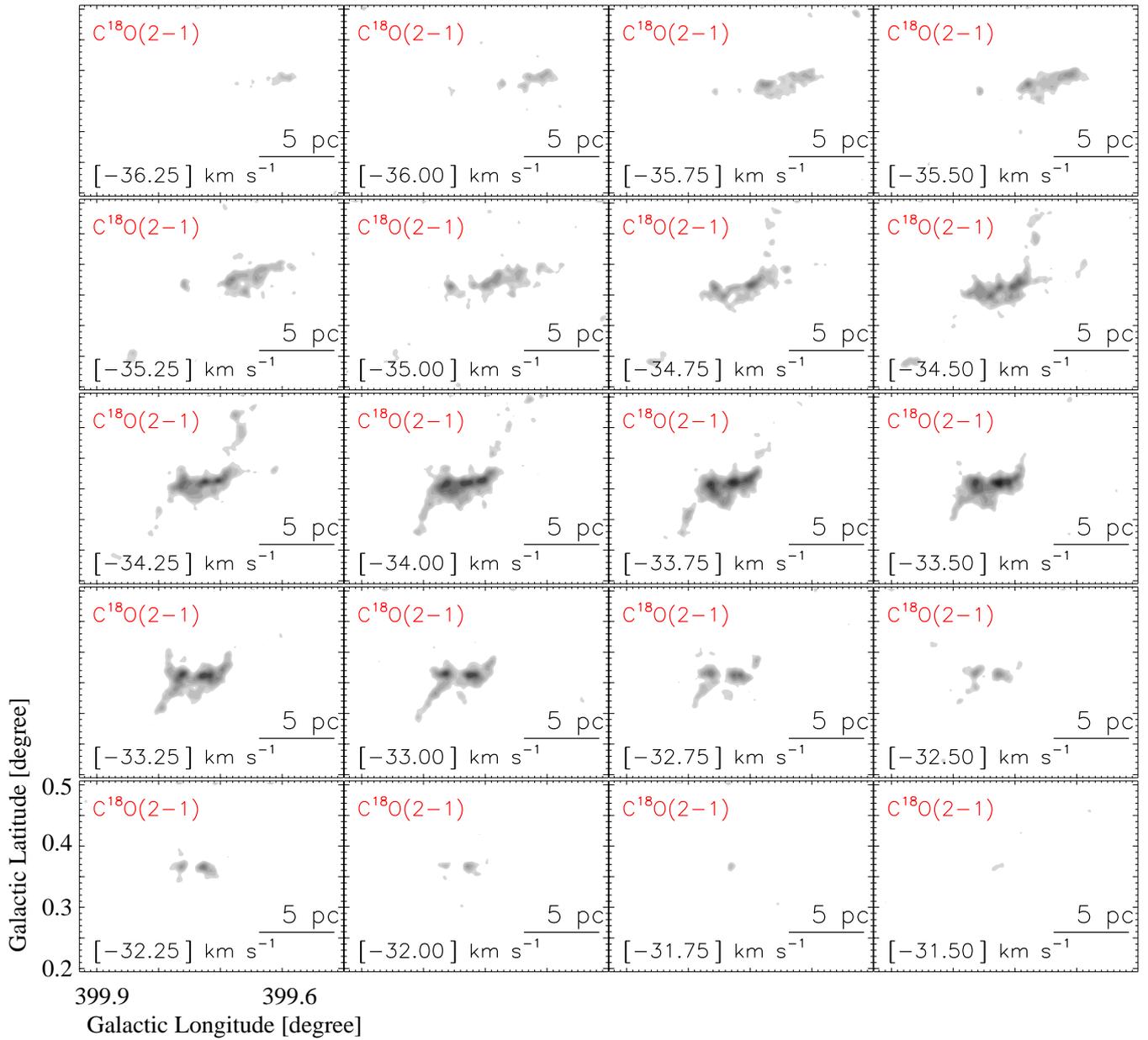}
\caption{Velocity channel maps of the SEDIGISM C$^{18}$O(2--1) emission. 
The velocity value (in km s$^{-1}$) is labeled in each panel. 
The contour levels are 1, 1.3, 1.5, 2, 2.5, 3, 3.5, 4, 4.3, 5, and 5.5 K. 
In each panel, the scale bar is the same as in Figure~\ref{fig1}a.}
\label{fig8}
\end{figure*}

In Figure~\ref{fig7}, to examine the gas motion in the filament, we display the velocity channel maps of $^{13}$CO(J = 2--1). In each panel of Figure~\ref{fig7}, we have marked two curves (i.e., ``xy" and ``xz"), allowing us to infer 
two cloud components (around $-$35.5 and $-$33.5 km s$^{-1}$) in the direction of the filament. 
In the velocity panel starting from $-$33.50 km s$^{-1}$ to $-$31.25 km s$^{-1}$, two molecular condensations are evident, 
and appear to host S1 and S2. In the panel at $-$33.25 km s$^{-1}$, the molecular condensation associated with S2 
is surrounded by filaments (see also the inset in Figure~\ref{fig1}a), favouring the presence of the HFS. 
This picture concerning HFS may not be entirely valid for the molecular condensation associated with S1 (see also Figure~\ref{fig1}b). 
These proposed arguments are also supported by the distribution of the C$^{18}$O(J = 2--1) emission (see the velocity channel maps of C$^{18}$O(J = 2--1) in Figure~\ref{fig8}). 

\begin{figure*}
\includegraphics[width=\textwidth]{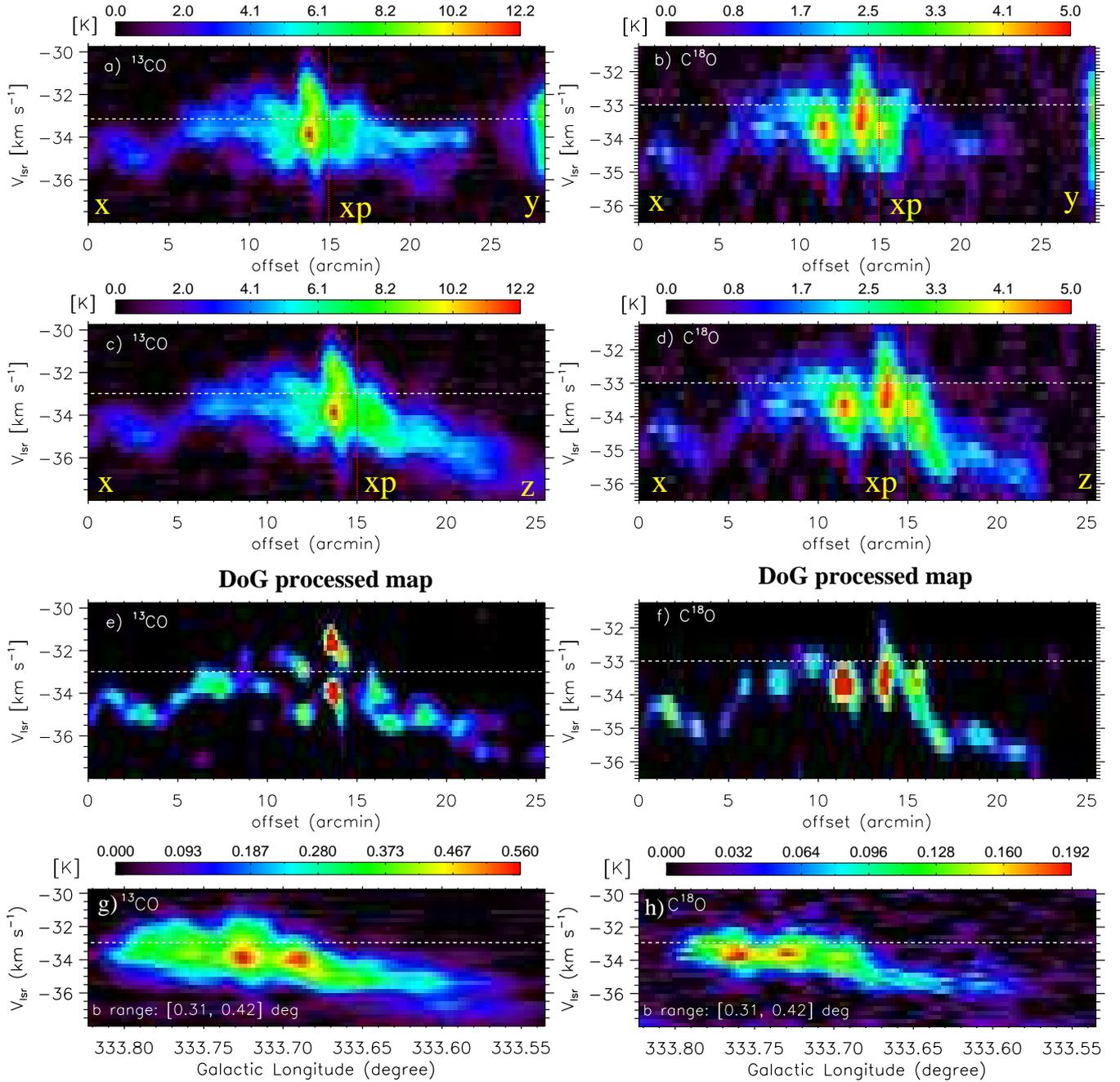}
\caption{Position-velocity diagram of a) $^{13}$CO(2--1); b) C$^{18}$O(J =2$-$1) along the curve ``xy" as 
highlighted in Figure~\ref{fig7}. 
Position-velocity diagram of c) $^{13}$CO(2--1); d) C$^{18}$O(J =2$-$1) along the curve ``xz" as 
highlighted in Figure~\ref{fig7}. e) The panel is the same as Figure~\ref{fig9}c, but is produced via the ``DoG" algorithm. 
f) The panel is the same as Figure~\ref{fig9}d, but is generated through the ``DoG" algorithm. 
Longitude-velocity diagram of g) $^{13}$CO(2--1); h) C$^{18}$O(2--1). 
The molecular emission is integrated over a latitude range (see a dotted-dashed box in 
Figure~\ref{fig2}b), which is indicated in Figures~\ref{fig9}g and~\ref{fig9}h. 
In each panel, a horizontal dashed line (in white) is shown at V$_\mathrm{lsr}$ = $-$33 km s$^{-1}$.}
\label{fig9}
\end{figure*}

In Figures~\ref{fig9}a and~\ref{fig9}b, we present the position-velocity diagrams of $^{13}$CO(J = 2--1) and C$^{18}$O(J = 2--1) along the curve ``xy" as indicated in Figure~\ref{fig7}, respectively. 
Figures~\ref{fig9}c and~\ref{fig9}d display the position-velocity diagrams of $^{13}$CO(J = 2--1) and C$^{18}$O(J = 2--1) along the curve ``xz" (see Figure~\ref{fig7}), respectively. 
In Figures~\ref{fig9}a--\ref{fig9}d, a vertical dotted line (in red) is marked, separating two zones. 
The velocity structure is the same toward the ``x-xp" zone in Figures~\ref{fig9}a and~\ref{fig9}c. 
Similarly, the velocity structure is identical toward the ``x-xp" zone in Figures~\ref{fig9}b and~\ref{fig9}d. 
To further examine velocity structures, we exposed the position-velocity diagrams of $^{13}$CO(J = 2--1) and C$^{18}$O(J = 2--1) to an edge detection algorithm \citep[i.e. Difference of Gaussian (DoG); see][]{gonzalez02,assirati14,dewangan17b}. These maps are presented in Figures~\ref{fig9}e and~\ref{fig9}f, 
and are displayed here only for a visual inspection purpose. Figures~\ref{fig9}e and~\ref{fig9}f are the same as Figures~\ref{fig9}c and~\ref{fig9}d, respectively, but are produced through the DoG algorithm. 
In Figures~\ref{fig9}e and~\ref{fig9}f, a noticeable velocity variation along the curve ``xz" is clearly evident. 
A significant velocity gradient is seen toward the  ``xp-z" zone compared to the ``x-xp" zone (see the velocity structure around the offset between 10$'$ to 18$'$ in Figures~\ref{fig9}c/\ref{fig9}e and~\ref{fig9}d/\ref{fig9}f). 
In Figures~\ref{fig9}g and~\ref{fig9}h, we show the Longitude-velocity diagrams of $^{13}$CO(J = 2--1) and C$^{18}$O(J = 2--1), respectively, where the molecular emission is integrated over a latitude range of [0$\degr$.31, 0$\degr$.42]. 

\citet{veena18} examined the profiles of HCO$^{+}$, H$^{13}$CO$^{+}$, HCN, and HNC toward the clump associated with S1, and reported signatures of infall. They also suggested the possibility of the presence of small-scale outflows toward S1. 
Most recently, \citet{yang22} examined the presence and absence of the $^{13}$CO outflows toward the ATLASGAL clumps using the SEDIGISM $^{13}$CO(J = 2--1) and C$^{18}$O(J = 2--1) line data. One can find the detailed steps for identifying the outflows toward ATLASGAL clumps in \citet{yang22}. In our selected target area, six ATLASGAL clumps are traced (see Figure~\ref{fig1}c), and the molecular outflow is reported toward only one ATLASGAL clump (i.e., AGAL333.724+00.364) hosting S1 \citep{yang22}. They only provided the velocity range of the $^{13}$CO(J = 2--1) blue wing-like velocity component at [$-$37.2, $-$35] km s$^{-1}$, and did not report the velocity range of the $^{13}$CO(J = 2--1) red wing-like velocity component. Previously, using the C$^{18}$O line data, \citet{shimoikura16} investigated the infalling motion with rotation of the cluster-forming clump in S235AB (distance $\sim$1.8 kpc). In this relation, they examined the position-velocity diagram along the major axis of the clump, which displays two well-defined peaks (at velocities $\sim$ $-$17.2 and $-$16.2 km s$^{-1}$) symmetrically located with respect to the center of the clump. The velocity separation of these two peaks is about 1 km s$^{-1}$. 

A visual inspection of Figures~\ref{fig9}a and~\ref{fig9}c also hints at the presence of the $^{13}$CO(J = 2--1) blue wing-like velocity component around an offset of 14$'$, which is spatially related to S1. But, we are unable to clearly assess the $^{13}$CO(J = 2--1) red wing-like velocity component toward S1, which may be due to the presence of another velocity component. One can keep in mind that the locations of S1 and S2 appear to be spatially located toward the overlapping areas of the two filamentary features traced in absorption (see arrows in Figure~\ref{fig1}b). However, at the same time, star formation signposts (i.e., infall and outflow) may not be ignored in the clump hosting S1 (see also Section~\ref{sec:ysos}). On the basis of the presence of two filamentary features, the velocity gradient seen in Figures~\ref{fig9}c and~\ref{fig9}d is unlikely to be explained by the inflow or outflow process. Therefore, our derived results through the analysis of the molecular line data favour the presence of two cloud components, which appear to intersect toward the areas hosting S1 and S2 or the circular regions, \#4--8 as indicated in Figure~\ref{fig3}c.

The implication of these findings is presented in Section~\ref{sec:disc}.
\subsection{Star formation activities in G333.73}
\label{sec:ysos}
As mentioned earlier, the central part of the filament hosts the bright infrared sources S1/IRAS 16164-4929 and S2/NIR31 associated with H\,{\sc ii} regions, which are powered by massive stars \citep{veena18}. Additionally, MIR 16 and NIR31 have been classified as YSOs \citep{veena18}. The Class~I protostar, MIR 16 was highlighted as an interesting source, and was detected at the center of the radio continuum emission at 1300 MHz \citep[see][for more details]{veena18}. 

To further assess the infrared-excess sources/YSOs and their distribution in the filament, we examined the photometric magnitudes of point-like sources in the H and K$_{s}$ bands from the VVV and 2MASS surveys. This work is benefited with high resolution of the VVV data. 
Using the NIR color-magnitude space (i.e. H$-$K$_{s}$/K$_{s}$; not shown here) and a color condition (H$-$K$_{s}$ $>$ 1.7), we find 595 candidate sources with 
infrared-excess emission. This color condition is selected through the study of a nearby control field. 
Following the work of \citet{dewangan15}, we produced the surface density map of these YSO candidates using a 10$''$ grid and 6 nearest-neighbor (NN) at a distance of 2.35 kpc. In Figures~\ref{fig11}a and~\ref{fig11}b, we overlaid the surface density contours of YSO candidates at [5, 10, 15] YSOs pc$^{-2}$ on the {\it Herschel} column density map and the $^{13}$CO(2--1) moment-0 map, respectively. On the basis of the surface density contours, the groups/clusters of YSO candidates are mainly evident in the direction of 12 circular regions (i.e., \#3--13 and \#15; see circles in Figure~\ref{fig11}b and also Figure~\ref{fig3}c). This particular analysis confirms the ongoing star formation activities (including massive stars) in the filament. 
\subsection{High-resolution NIR and mm maps}
\label{sec:morphss}
\begin{figure*}
\includegraphics[width=\textwidth]{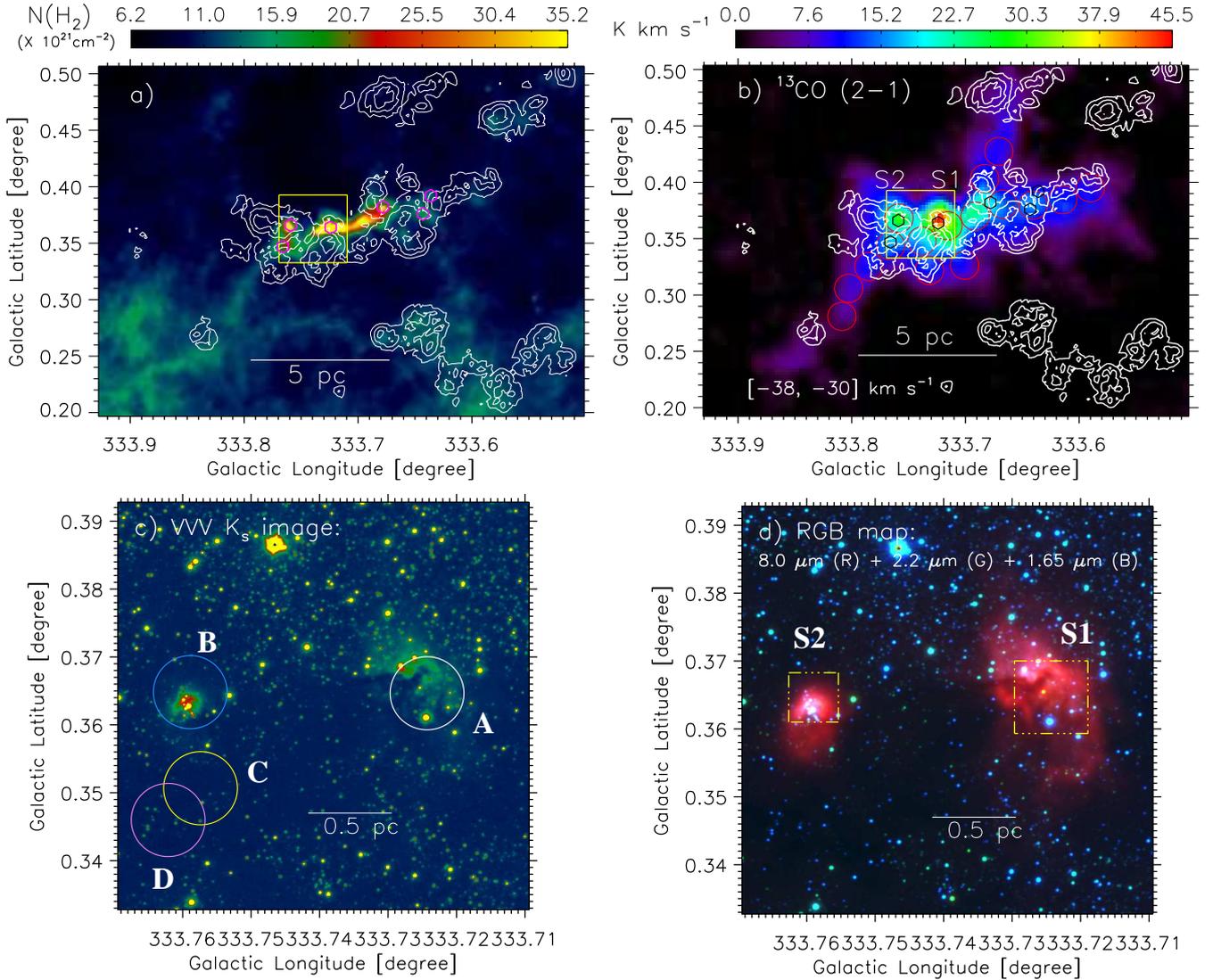}
\caption{Overlay of surface density contours of YSOs/infrared-excess sources on a) the {\it Herschel} column density map; 
b) the $^{13}$CO(2--1) moment-0 map. 
c) The panel displays the VVV K$_{s}$ image containing two MIR 
sources S1 and S2 (see a solid box in Figures~\ref{fig11}a and~\ref{fig11}b). 
Big circles highlight the areas covered by ALMA band-6 observations. 
d) The panel presents a three color-composite map of an area hosting 
S1 and S2 (see a solid box in Figures~\ref{fig11}a and~\ref{fig11}b). 
The color-composite map is produced using three images: 8.0 $\mu$m (red), K$_{s}$ (green), and H (blue).
In panels ``a" and ``b", the surface density contours (in white) are shown 
with the levels of 5, 8, and 15 YSOs pc$^{-2}$, and the positions of the ATLASGAL dust continuum 
clumps at 870 $\mu$m \citep[from][]{urquhart18} are also marked by hexagons. 
In panel ``b", 16 circles are the same as presented in Figure~\ref{fig3}c.} 
\label{fig11}
\end{figure*}

Figures~\ref{fig11}c and~\ref{fig11}d present the VVV K$_{s}$ image and a three color-composite map (RGB: 8.0 $\mu$m (in red), K$_{s}$ (in green), and H (in blue) images) of an area hosting S1 and S2 (see a solid box in Figures~\ref{fig11}a and~\ref{fig11}b).
In the direction of S2, a compact MIR emission surrounding several point-like sources (including NIR31) is evident in the infrared images. An extended MIR emission hosting the embedded source MIR 16 at its center is found in the direction of S1. 

\begin{figure*}
\includegraphics[width=\textwidth]{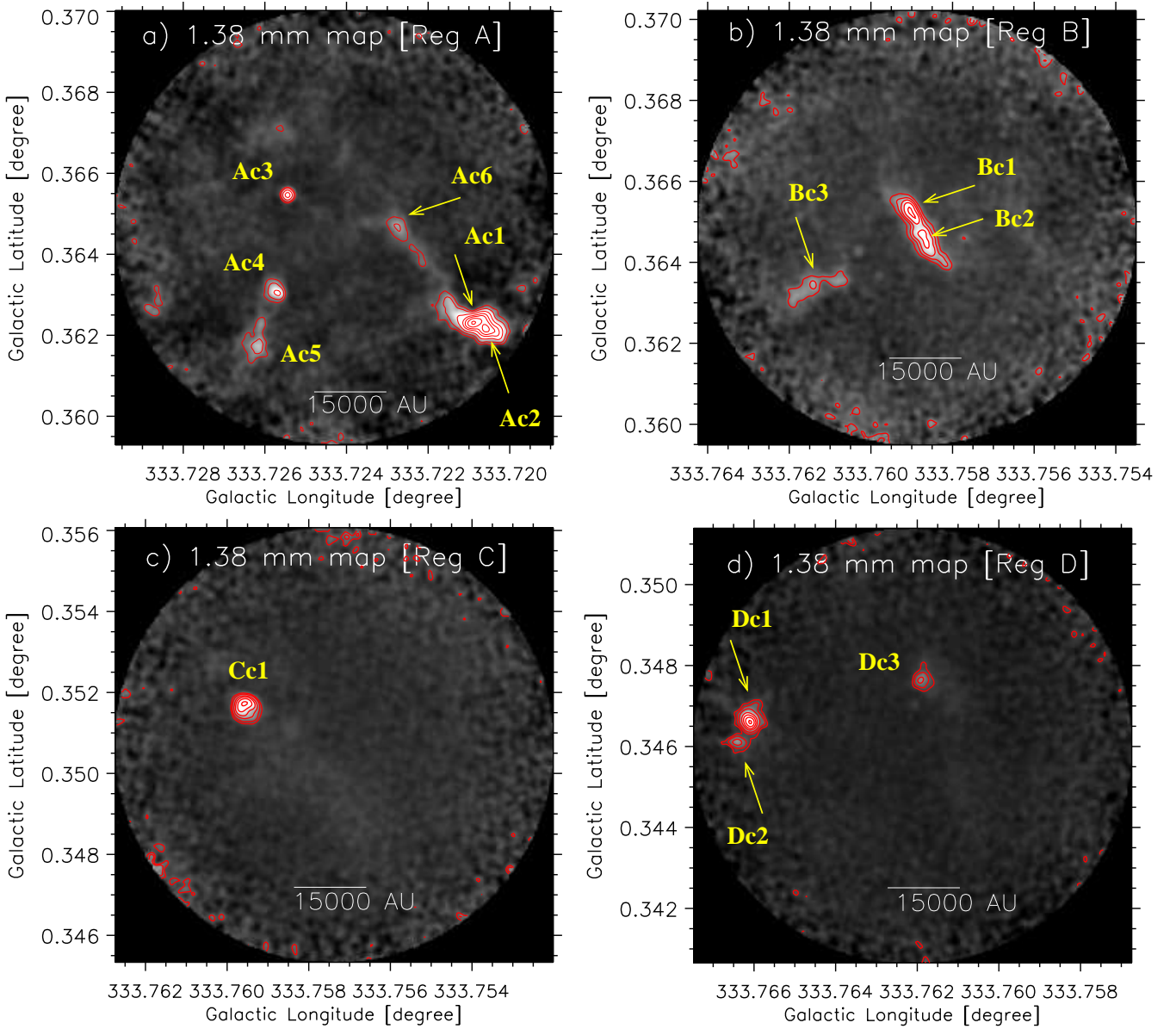}
\caption{ALMA 1.38 mm continuum map of a) Reg ``A"; b) Reg ``B"; c) Reg ``C"; 
d) Reg ``D" (see big circles in Figure~\ref{fig11}c). 
In panel ``a", the 1.38 mm continuum contours are displayed with the levels of 
(0.05, 0.1, 0.2, 0.35, 0.5, 0.7, 0.95) $\times$ 20.17 mJy beam$^{-1}$. 
In panel ``b", the 1.38 mm continuum contours are shown with the levels of 
(0.2, 0.35, 0.5, 0.7, 0.95) $\times$ 4.14 mJy beam$^{-1}$. 
In panel ``c", the 1.38 mm continuum contours are presented with the levels of 
(0.12, 0.2, 0.35, 0.5, 0.7, 0.95) $\times$ 6.1 mJy beam$^{-1}$. 
In panel ``d", the 1.38 mm continuum contours are drawn with the levels of 
(0.07, 0.12, 0.2, 0.35, 0.5, 0.7, 0.95) $\times$ 12.24 mJy beam$^{-1}$.}
\label{fig12}
\end{figure*}

At least four pointings (i.e., ``Reg A -- Reg D") are found toward the central part of the filament, where the ALMA 
continuum maps at 1.38~mm are available (see big circles in Figure~\ref{fig11}c). One can notice that the ALMA pointings ``Reg A" and ``Reg B" are found toward S1 and S2, respectively. 
The ALMA 1.38~mm continuum map and contours of Reg A, Reg B, Reg C, and Reg D are presented in Figures~\ref{fig12}a,~\ref{fig12}b,~\ref{fig12}c, and~\ref{fig12}d, respectively. To find out the continuum sources in each ALMA map at 1.38 mm \citep[see also][]{dewangan21g25}, we used the {\it clumpfind} IDL program \citep{williams94}, enabling us to obtain the total flux, the FWHM not corrected for beam size for the x-axis (i.e., FWHM$_{x}$), and for the y-axis (i.e., FWHM$_{y}$). In Reg A, Reg B, Reg C, and Reg D, we find six (i.e., Ac1--Ac6), 
three (i.e., Bc1--Bc3), one (i.e., Cc1), and three (i.e., Dc1--Dc3) continuum sources/cores, respectively. 
We have listed the fluxes, deconvolved FWHM$_{x}$ \& FWHM$_{y}$, and masses of all the continuum sources in Table~\ref{tab3}. 

Concerning the mass calculation of each core, we adopted the following equation \citep{hildebrand83}: 
\begin{equation}
M \, = \, \frac{D^2 \, F_\nu \, R_t}{B_\nu(T_d) \, \kappa_\nu}
\end{equation} 
\noindent where $F_\nu$ is the total integrated flux (in Jy), 
$D$ is the distance (in kpc), $R_t$ is the gas-to-dust mass ratio, 
$B_\nu$ is the Planck function for a dust temperature $T_d$, 
and $\kappa_\nu$ is the dust absorption coefficient. 
This work utilizes $\kappa_\nu$ = 0.9\,cm$^2$\,g$^{-1}$ at 1.3 mm \citep{ossenkopf94}, 
$T_d$ = 22~K (see circular regions highlighted by a dagger in Table~\ref{tab2}), and $D$ = 2.35 kpc. 
All the selected continuum sources are revealed as low-mass cores (see Table~\ref{tab3}). 
We may suggest the uncertainty in the mass estimation to be typically $\sim$20\% and at maximum $\sim$50\%, which could be due to uncertainties in the adopted opacity, dust temperature, and measured flux.  

\begin{figure*}
\includegraphics[width=\textwidth]{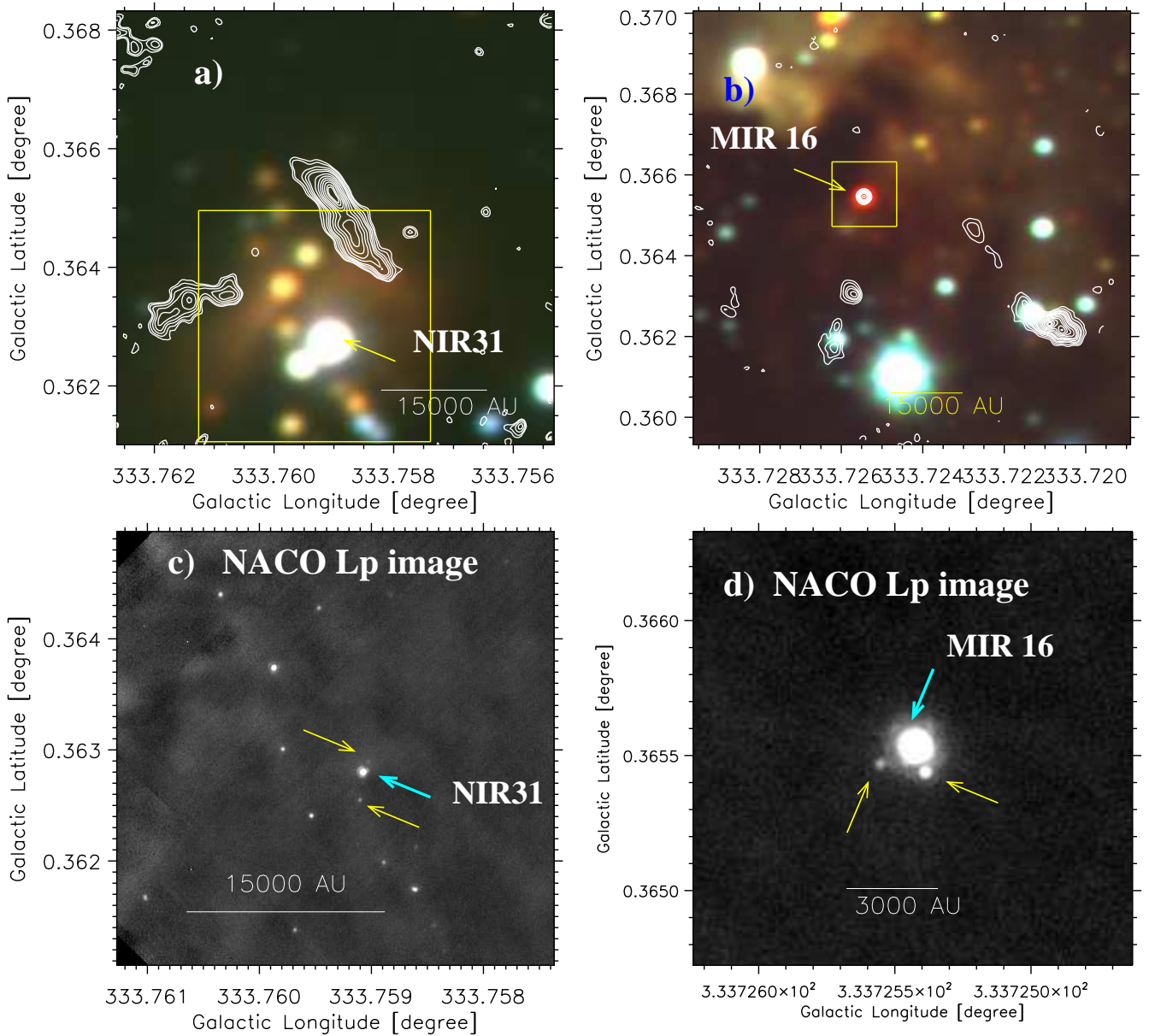}
\caption{a) Overlay of the ALMA 1.38 mm continuum emission contours on a three color-composite map of an area hosting S2 (see a dotted-dashed box in Figure~\ref{fig11}d). 
The 1.38 mm continuum contours are displayed with the levels of 
(0.12, 0.15, 0.2, 0.25, 0.3, 0.4, 0.5, 0.6, 0.7, 0.8, 0.9, 0.95) $\times$ 4.13 mJy beam$^{-1}$.
b) Overlay of the ALMA 1.38 mm continuum emission contours on a three color-composite map of an area hosting S1 (see a dotted-dashed box in Figure~\ref{fig11}d). 
The 1.38 mm continuum contours are displayed with the levels of 
(0.05, 0.08, 0.12, 0.15, 0.2, 0.25, 0.3, 0.4, 0.5, 0.6, 0.7, 0.8, 0.9, 0.95) $\times$ 20.26 mJy beam$^{-1}$.
c) The VLT/NACO adaptive-optics L$^{\prime}$-band image of an area highlighted by a solid box in Figure~\ref{fig13}a. 
d) The VLT/NACO adaptive-optics L$^{\prime}$-band image of an area highlighted by a solid box in Figure~\ref{fig13}b. 
In panels ``a" and ``b", the color-composite map is produced using three VVV images: K$_{s}$ (red), H (green), and J (blue). 
In panels ``c" and ``d", arrows highlight point-like sources seen below the 4000 AU scale.} 
\label{fig13}
\end{figure*}

Figures~\ref{fig13}a and~\ref{fig13}b present the overlay of the ALMA 1.38 mm continuum emission contours on 
a three color-composite map using VVV (RGB: K$_{s}$ (red), H (green), and J (blue)) images of an area hosting S2 and S1, respectively (see dotted-dashed boxes in Figure~\ref{fig11}d). 
The 1.38 mm continuum emission is not observed toward the YSO NIR31, which was proposed as an exciting source of the H\,{\sc ii} region. 
Interestingly, in the ALMA map, the compact continuum emission is detected toward the Class~I YSO, MIR 16. In Figures~\ref{fig13}c and~\ref{fig13}d, we have presented the VLT/NACO L$^{\prime}$-band images (resolution $\sim$0\farcs1) toward the YSOs NIR31 and MIR 16, respectively (see a solid box in Figures~\ref{fig13}a and~\ref{fig13}b), allowing us to infer inner structures around the objects. Below the 4000 AU scale, at least three point-like sources are seen toward both the YSOs (see arrows in Figures~\ref{fig13}c and~\ref{fig13}d), but 
no small-scale features are found around these YSOs. 
\section{Discussion}
\label{sec:disc}
Despite the previously reported multi-wavelength study of the G333.73 filament, the birth process of massive stars is yet to be explored. In this context, this paper deals with a multi-scale and multi-wavelength study of the G333.73 filament. In Section~\ref{ssec:mmap}, we find the presence of nonthermal (or 
turbulent) motions in the filament, and its central part contains earlier studied two H\,{\sc ii} regions.
The filament is also associated with star formation activities (see Section~\ref{sec:ysos}). 
From Figure~\ref{fig1}b, we find the presence of two filamentary structures in the G333.73 filament. 
A HFS associated with S2 is investigated in the filament. 
In the direction of S1 and S2, low-mass cores are detected using the ALMA continuum map (see Section~\ref{sec:morphss}). 

The observed HFS associated with S2 indirectly reflects the inflow material toward the central hub from a large scale of 1--10 pc, which can be channeled through the molecular cloud filaments \citep[e.g.,][]{trevino19}.
According to an evolutionary scheme for MSF associated with the HFSs \citep[see][]{tige_2017,motte18}, low mass stellar embryos will be responsible for infrared bright massive protostars. These low mass stellar embryos are expected to gain the material via gravitationally-driven inflows from the parental massive dense cores/clumps. 
Hence, the presence of low-mass cores toward the HFS associated with S2 gives an impression that the cores producing massive stars may not gain the entire mass prior to core collapse, but instead, these entities gain mass simultaneously \citep[e.g.,][]{zhang09,wang11,wang14,sanhueza19,svoboda19,dewangna21w42}. Our analysis enables us to predict that in future, the central part of the G333.73 filament will form new O-type stars.

The analysis of the $^{13}$CO(J = 2--1) and C$^{18}$O(J = 2--1) emission reveals a new picture in the G333.73 filament, which is the existence of the two velocity components (around $-$35.5 and $-$33.5 km s$^{-1}$) in the direction of the filament (see Section~\ref{ssec:mmap}). 
This result and the two filamentary structures seen in the {\it Spitzer} 8.0 $\mu$m image together favour the presence of 
two filaments or filamentary clouds at different velocities toward our target site. It seems that the MSF and the clustering of YSO candidates are concentrated toward the central part of the filament where the two filamentary clouds having a velocity separation of about 2 km s$^{-1}$ intersect. 
These observational findings seem to favour the theory of CCC or converging flows in the G333.73 filament. 
 
Concerning the CCC process, numerical simulations of a head-on collision of two dissimilar clouds were carried out by 
\citet{habe92}. They found gravitationally unstable cores/clumps at the interface of the clouds in the simulations, 
which were produced due to the effect of the compression of the colliding clouds. 
Clusters of YSOs and massive stars are strongly expected at the intersection of the colliding clouds or the shock-compressed interface layer. In relation to the scenario of converging flows, it has been proposed that molecular clouds can be the result of the convergence of streams of neutral gas \citep{ballesteros99,vazqez07,heitsch08}. With time, there is possibility of the merging/converging/collision of the filaments of molecular gas, which can lead the formation of cores and stars \citep[e.g.,][]{galvan10,beuther20,dewangan20w33}.
In general, CCC may be treated as the high-density form of colliding gas flows in a bigger manner, which is initiated in the low-density medium \citep{beuther20}. 

Concerning head-on collision of two clouds, the smoothed particle hydrodynamics simulations also predict the existence of a configuration of filaments (e.g., hub or spokes systems) through the collision event \citep{balfour15}. In the simulations, a shock-compressed layer is generated by the colliding clouds, and fragments into filaments. Such filaments then can be responsible for a network like a spider's web. The latest review article on CCC by \citet{fukui21} also favours the origin of the hub filaments with their complex morphology in the collision process. Together, one of the strongest predictions of these scenarios is the existence of a shock-compressed interface layer by the colliding clouds/flows, where HFSs, massive stars, and clusters of YSOs can be originated.

Following the equation given in \citet{henshaw13}, one can infer the time-scale for which the material is accumulated at 
the collision zones or the collision time-scale, which is defined as
\begin{equation}
t_{\rm accum} = 2.0\,\bigg(\frac{l_{\rm fcs}}{0.5\,{\rm pc}} \bigg) \bigg(\frac{v_{\rm
rel}}{5{\rm \,km\,s^{-1}}}\bigg)^{-1}\bigg(\frac{n_{\rm pstc}/n_{\rm
prec}}{10}\bigg)\,{\rm Myr} 
\end{equation}
where, $n_{\rm prec}$ and $n_{\rm pstc}$ are the average densities of the pre-collision and post-collision region, respectively.
$l_{\rm fcs}$ is the collision length-scale, while ${v_{\rm rel}}$ is the observed relative velocity. 
In this study, the spatial extent of the overlapping regions of the two clouds is $\sim$2 pc, which have a velocity separation of $\sim$2 km s$^{-1}$. Due to unknown viewing angle of the collision, we assume a typical viewing angle of 45$\degr$, resulting  
$l_{\rm fcs}$ $\sim$2.8 pc ($\sim$2 pc/sin(45$\degr$)) and ${v_{\rm rel}}$ $\sim$2.8 km s$^{-1}$ ($\sim$2 km s$^{-1}$/cos(45$\degr$)). Reliable estimates of $n_{\rm prec}$ and $n_{\rm pstc}$ values are beyond the scope of this paper. But, in general, the collision favours $n_{\rm pstc}$ $>$ $n_{\rm prec}$. Therefore, a range of collision timescale is determined to be $\sim$2--20 Myr for a given range of the mean density ratio of 1--10. This analysis indicates a representative time-scale of the collision of two clouds to be $\sim$2 Myr. Earlier, the dynamical ages of H\,{\sc ii} regions associated with S1 and S2 were reported to be 0.2 and 0.01 Myr respectively \citep{veena18}, which are much younger than $t_{\rm accum}$. 

In the {\it Herschel} column density map, the peak column density of 5.8--7.8 $\times$ 10$^{22}$ cm$^{-2}$ is found in the direction of the overlapping regions of the two clouds (see locations around S1 and S2 in Figure~\ref{fig2}b), where star formation activities (including massive stars) are evident (see Figure~\ref{fig11}). \citet{fukui21b} carried out magnetohydrodynamic (MHD) simulations of colliding molecular flows, and studied the O-type star formation by exploring column density in the filaments. 
They reported the dependencies in column density for the formation of O-type stars in isolation and/or in a cluster. 
They suggested that 10 O stars (or a single O star) can be produced for total column density of 
10$^{23}$ (10$^{22}$) cm$^{-2}$ \citep[see also][]{inoue18,liow20}. 

In the direction of N159E-Papillon Nebula and N159W-South in the Large Magellanic Cloud (distance $\sim$50 kpc), 
the presence of massive stars and HFSs has been reported \citep[e.g.,][and references therein]{fukui19ex,tokuda19ex}. 
N159E-Papillon cloud is located $\sim$50 pc away from the N159W-South region in projection \citep[see Figure~7 in][]{fukui19ex}. 
Using the ALMA $^{13}$CO (2--1) line data, velocity structures toward both the regions were examined.  
A noticeable velocity gradient was observed toward N159E-Papillon cloud \citep[see Figure~3 in][]{fukui19ex} and  
the N159W-South \citep[see Figure~5 in][]{tokuda19ex}. In connection to the formation of hub-filaments and massive stars, using the ALMA observations, 
\citet{fukui19ex} and \citet{tokuda19ex} suggested the applicability of a scenario of the large-scale colliding flow in the N159E-Papillon Nebula and N159W-South. CCC was proposed to be operated around 2 Myr ago in N159E-Papillon \citep[see Figure~7 in][]{fukui19ex}. 
This proposed scenario was in agreement with the MHD numerical simulations 
of \citet{inoue18}. Such collision event produces the shock-compressed layer where massive filaments with 
massive stars at their vertex are expected \citep[see][for more details]{inoue18}. 
In this paper, we also find a significant velocity gradient toward the ``xp-z" zone (see the velocity structure around the offset between 10$'$ to 18$'$ in Figures~\ref{fig9}c/\ref{fig9}e and~\ref{fig9}d/\ref{fig9}f). The observed velocity structure and the collisional time-scale in our target site appear to be very similar as reported toward N159E-Papillon Nebula and N159W-South. 

Recently, \citet{wu18} studied three-dimensional turbulent MHD simulations of non-colliding and colliding giant molecular clouds (GMCs). They found collisions between GMCs can produce higher turbulence within dense gas structures, which has been observationally reported in dense IRDCs. On the other hand, in the case of non-colliding GMCs, one may expect turbulent decay with much less energetic high-density structures \citep[see also][]{sakre21}. 
Hence, it is likely that the colliding filaments may be the potential driver of the nonthermal (or turbulent) motions in our target IRDC. 

Overall, in the G333.73 filament, star formation activities (including massive stars) and the HFS seem to be explained by the scenario of the converging/colliding flows from two different velocity components. 
\section{Summary and Conclusions}
\label{sec:conc}
To probe the ongoing star formation process in the G333.73 filament, an observational study is carried out in this paper. 
The paper has utilized a multi-scale and multi-wavelength approach, allowing to examine the cloud to core scale physical environments of the G333.73 filament. 
The SEDIGISM $^{13}$CO(J =2$-$1) and C$^{18}$O(J =2$-$1) line data are examined to study 
the structure and kinematics of the molecular gas in the filament. 
In addition to the existing large scale infrared and sub-mm continuum surveys 
(i.e., {\it Spitzer}-GLIMPSE, WISE, {\it Herschel}, ATLASGAL), the ALMA 1.38 mm dust continuum 
map (resolution $\sim$0\farcs8), the VLT/NACO adaptive-optics L$^{\prime}$-band 
images (resolution $\sim$0\farcs1), and the VVV NIR data (resolution $\sim$0\farcs8) are explored. 

At least two filamentary structures (length $>$ 6 pc) and a HFS associated with S2 are identified in absorption using the {\it Spitzer} 8.0 $\mu$m image. 
The $^{13}$CO(J = 2--1) and C$^{18}$O(J = 2--1) emissions show two velocity components (around $-$35.5 and $-$33.5 km s$^{-1}$) in the direction of the G333.73 filament, supporting the presence of two filaments or filamentary clouds at different velocities. Nonthermal (or turbulent) motions in the filament are investigated using the $^{13}$CO and C$^{18}$O line data. 
Based on the distribution of YSO candidates, we trace star formation activities in the filament, which also hosts 
previously known two H\,{\sc ii} regions at its center. 
In the direction of these H\,{\sc ii} regions associated with S1 and S2, low-mass cores are identified using the ALMA continuum map. 
The VLT/NACO adaptive-optics L$^{\prime}$-band images show the presence of at least three point-like sources and the absence 
of small-scale features in the inner 4000 AU around two previously known YSOs NIR31 and MIR 16.  
The locations of H\,{\sc ii} regions and groups of infrared-excess sources are evident toward the central part of the G333.73 filament, where the two filamentary clouds intersect. 

Our outcomes appear to support the scenario of CCC or converging flows in the G333.73 filament, which may also explain the observed HFS. We also propose that in the direction of the HFS associated with S2, 
new O-type stars may form in future. 
\section*{Acknowledgments}
We are grateful to the anonymous reviewer for the constructive comments and 
suggestions, which greatly improved the scientific contents of the paper. 
The research work at Physical Research Laboratory is funded by the Department of Space, Government of India. 
This work is based [in part] on observations made with the {\it Spitzer} Space Telescope, which is operated by the Jet Propulsion Laboratory, California Institute of Technology under a contract with NASA. This publication is based on data acquired with the Atacama Pathfinder Experiment (APEX) under programmes 092.F-9315 and 193.C-0584. APEX is a collaboration among the Max-Planck-Institut fur Radioastronomie, the European Southern Observatory, and the Onsala Space Observatory. The processed data products are available from the SEDIGISM survey database located at https://sedigism.mpifr-bonn.mpg.de/index.html, which was constructed by James Urquhart and hosted by the Max Planck Institute for Radio Astronomy. 
This paper makes use of the following ALMA archive data: ADS/JAO.ALMA\#2016.1.00191.S. ALMA is a partnership of ESO (representing its member states), NSF (USA) and NINS (Japan), together with NRC (Canada), MOST and ASIAA (Taiwan), and KASI (Republic of Korea), in cooperation with the Republic of Chile. The Joint ALMA Observatory is operated by ESO, AUI/NRAO and NAOJ. 
\subsection*{Data availability}
The ALMA continuum data underlying this article are available from the publicly accessible JVO ALMA FITS archive\footnote[1]{http://jvo.nao.ac.jp/portal/alma/archive.do/}.
The SEDIGISM molecular line data underlying this article are available from the publicly accessible website\footnote[2]{https://sedigism.mpifr-bonn.mpg.de/index.html}. The {\it Herschel} temperature and column density maps underlying this article are available from the publicly accessible website\footnote[3]{http://www.astro.cardiff.ac.uk/research/ViaLactea/}.
The {\it Herschel}, {\it Spitzer}, and 2MASS data underlying this article are available from the publicly accessible NASA/IPAC infrared science archive\footnote[4]{https://irsa.ipac.caltech.edu/frontpage/}.
The ESO VLT/NACO data underlying this article are available from the publicly accessible ESO website\footnote[5]{https://archive.eso.org/eso/eso\_archive\_main.html}. 
The VVV data underlying this article are available from the publicly accessible website\footnote[6]{http://horus.roe.ac.uk/vsa/vvvGuide.html}.
\begin{table*}
\scriptsize
\setlength{\tabcolsep}{0.1in}
\centering
\caption{List of different surveys studied in this paper.}
\label{tab1}
\begin{tabular}{lcccr}
\hline 
  Survey  &  Wavelength/Frequency/line(s)       &  Resolution ($\arcsec$)        &  Reference \\   
\hline
\hline 
SEDIGISM &  $^{13}$CO, C$^{18}$O (J = 2--1) & $\sim$30        &\citet{schuller17}\\
APEX Telescope Large Area Survey of the Galaxy (ATLASGAL)                 &870 $\mu$m                     & $\sim$19.2        &\citet{schuller09}\\
{\it Herschel} Infrared Galactic Plane Survey (Hi-GAL)                              &70--500 $\mu$m                     & $\sim$6--37         &\citet{molinari10}\\
Wide Field Infrared Survey Explorer (WISE) & 12 $\mu$m                   & $\sim$6           &\citet{wright10}\\ 
{\it Spitzer} Galactic Legacy Infrared Mid-Plane Survey Extraordinaire (GLIMPSE)       &3.6--8.0  $\mu$m                   & $\sim$2           &\citet{benjamin03}\\
Vista Variables in the V\'{\i}a L\'{\i}actea (VVV)                                  & 1.25--2.2 $\mu$m                            & $\sim$0.8             & \citet{minniti10}\\ 
Two Micron All Sky Survey (2MASS)                                                   & 1.25--2.2 $\mu$m                            & $\sim$2.5             & \citet{skrutskie06}\\
\hline          
\end{tabular}
\end{table*}

\begin{table*}
\setlength{\tabcolsep}{0.1in}
\centering
\caption{Table provides physical parameters toward 16 circular regions as marked in Figure~\ref{fig3}c. Table contains 
(region id, positions, average dust temperature (T$_{d}$), C$^{18}$O line velocity (V$_{lsr}$), C$^{18}$O line 
width ($\Delta V$), sound speed ($a_{s}$), non-thermal velocity dispersion ($\sigma_{NT}$), Mach number, and 
ratio of thermal to non-thermal gas pressure (R$_{p}$ = ${a^2_{s}}/{\sigma^2_{NT}}$) (see also Figure~\ref{fig5}). 
The selected regions located near S1 and S2 are highlighted by superscript ``$\dagger$".}
\label{tab2}
\begin{tabular}{lccccccccccccc}
\hline 										    	        			      
  Region     &   Longitude 	   &      Latitude    &     T$_{d}$    &       V$_{lsr}$    &       $\Delta V$          &     $a_{s}$	        &   $\sigma_{NT}$& Mach number  & R$_{p}$      \\ 
             &   [degree]	   &    [degree]      &     (K)          & 	(km s$^{-1}$) &      (km s$^{-1}$)        &   (km s$^{-1}$)     &   (km s$^{-1}$)& &  (${a^2_{s}}/{\sigma^2_{NT}}$) \\ 
\hline 
  1            &    333.809   &    0.281  &   19.4  &	$-$33.9   &  0.4   &  0.26   &   0.18	&   0.7  &    2.1 \\
  2            &    333.803   &    0.306  &   19.4  &	$-$33.6   &  0.6   &  0.26   &   0.26	&   1.0  &    1.0 \\
  3            &    333.785   &    0.325  &   19.8  &	$-$33.4   &  0.7   &  0.26   &   0.29	&   1.1  &    0.8 \\
  4$^{\dagger}$&    333.766   &    0.344  &   20.1  &   $-$33.7   &  0.7   &  0.26   &   0.31	&   1.2  &    0.7 \\
  5$^{\dagger}$&    333.758   &    0.368  &   21.0  &   $-$33.7   &  1.1   &  0.27   &   0.48	&   1.8  &    0.3 \\
  6$^{\dagger}$&    333.738   &    0.348  &   20.4  &   $-$33.9   &  0.8   &  0.26   &   0.31	&   1.2  &    0.7 \\
  7$^{\dagger}$&    333.715   &    0.361  &   21.5  &   $-$33.6   &  0.9   &  0.27   &   0.39	&   1.4  &    0.5 \\
  8$^{\dagger}$&    333.691   &    0.370  &   18.2  &   $-$34.2   &  1.0   &  0.25   &   0.41	&   1.7  &    0.3 \\
  9            &    333.667   &    0.380  &   18.8  &	$-$35.1   &  0.7   &  0.25   &   0.29	&   1.2  &    0.7 \\
  10           &    333.640   &    0.380  &   18.9  &	$-$35.3   &  0.7   &  0.25   &   0.27	&   1.1  &    0.9 \\
  11           &    333.614   &    0.385  &   19.5  &	$-$35.7   &  0.6   &  0.26   &   0.24	&   0.9  &    1.2 \\
  12           &    333.590   &    0.395  &   19.5  &	$-$35.6   &  0.8   &  0.26   &   0.32	&   1.2  &    0.7 \\
  13           &    333.731   &    0.322  &   19.7  &	$-$33.8   &  0.8   &  0.26   &   0.34	&   1.3  &    0.6 \\
  14           &    333.701   &    0.326  &   19.7  &	$-$34.0   &  1.0   &  0.26   &   0.42	&   1.6  &    0.4 \\
  15           &    333.683   &    0.403  &   19.6  &	$-$34.1   &  0.8   &  0.26   &   0.34	&   1.3  &    0.6 \\
  16           &    333.671   &    0.427  &   19.7  &	$-$34.4   &  0.5   &  0.26   &   0.20	&   0.8  &    1.7 \\
\hline          		
\end{tabular}			
\end{table*}			

\begin{table*}
\setlength{\tabcolsep}{0.1in}
\centering
\caption{Physical parameters of ALMA 1.38 mm continuum sources (see Figure~\ref{fig12}). 
Table lists names, positions, flux densities, deconvolved FWHM$_{x}$ \& FWHM$_{y}$, and masses of 
the continuum sources.} 
\label{tab3}
\begin{tabular}{lccccccccccccc}
\hline 										    	        			      
  Name           &  Longitude         & Latitude         & Total Flux &FWHM$_{x}$ $\times$ FWHM$_{y}$ & Mass (M$_{\odot}$)   \\ 
                 &  (degree)         &  (degree)         &     (mJy)  & ($''$ $\times$ $''$)          &  at $T_D$ = 22 K        \\ 
\hline 
Ac1 & 333.721 &   0.362   &	 65.09  &    2.3 $\times$ 1.6 &   6.9	\\
Ac2 & 333.721 &   0.362   &	 57.70  &    1.5 $\times$ 1.4 &   6.1	\\
Ac3 & 333.725 &   0.365   &	  7.96  &    0.7 $\times$ 0.7 &   0.8	\\
Ac4 & 333.726 &   0.363   &	 10.23  &    1.0 $\times$ 1.1 &   1.1	\\
Ac5 & 333.726 &   0.362   &	 11.94  &    1.1 $\times$ 2.2 &   1.3	\\
Ac6 & 333.723 &   0.365   &	 05.97  &    1.0 $\times$ 1.2 &   0.6	\\
Bc1 & 333.759 &   0.365   &	 16.89  &    1.4 $\times$ 1.6 &   1.8	\\
Bc2 & 333.759 &   0.365   &	 23.36  &    1.9 $\times$ 2.4 &   2.5	\\
Bc3 & 333.761 &   0.363   &	  9.85  &    3.4 $\times$ 1.6 &   1.0	\\
Cc1 & 333.760 &   0.352   &	 22.99  &    1.4 $\times$ 1.4 &   2.4	\\
Dc1 & 333.766 &   0.347   &	 30.98  &    1.2 $\times$ 1.4 &   3.3	\\
Dc2 & 333.766 &   0.346   &	  6.54  &    1.2 $\times$ 0.9 &   0.7	\\
Dc3 & 333.762 &   0.348   &	  6.54  &    1.0 $\times$ 1.2 &   0.7	\\
\hline          		
\end{tabular}			
\end{table*}			


\bibliographystyle{mnras}
\bibliography{reference} 



\end{document}